\documentclass[twocolumn]{aa}  

\usepackage{graphicx}
\usepackage{txfonts}
\usepackage{lipsum}
\usepackage{subcaption}         
\usepackage{lscape}             
\usepackage{placeins}           
                                
\begin{document}

   \title{Limits on a Host Star around a Saturn-mass Free-floating Planet Candidate KMT-2024-BLG-0792/OGLE-2024-BLG-0516}
   \titlerunning{Limits on a Host Star around KMT-2024-BLG-0792/OGLE-2024-BLG-0516}
   \authorrunning{Mr\'oz et al.}


%
%
%

   \author{Przemek Mr\'oz\inst{1}\corrauth{pmroz@astrouw.edu.pl}
        \and Klara Piotrowska\inst{2}\email{kpiotrowska@clipper.ens.psl.eu}
        \and Antoine M\'erand\inst{3}\email{amerand@eso.org}
        \and Subo Dong\inst{4,5,6}\email{dongsubo@pku.edu.cn}
        \and Zexuan Wu\inst{4,5}\email{wuzexuan@pku.edu.cn}
        }

   \institute{Astronomical Observatory, University of Warsaw, Al.~Ujazdowskie 4, 00-478 Warszawa, Poland 
   \and D\'epartement de Physique, \'Ecole Normale Sup\'erieure, Paris, France 
   \and European Southern Observatory, Karl-Schwarzschild-Stra\ss{}e 2, D-85748 Garching, Germany 
   \and Department of Astronomy, School of Physics, Peking University, 5 Yiheyuan Road, Haidian District, Beijing 100871, People's Republic of China 
   \and Kavli Institute of Astronomy and Astrophysics, Peking University, 5 Yiheyuan Road, Haidian District, Beijing 100871, People's Republic of China 
   \and National Astronomical Observatories, Chinese Academy of Science, 20A Datun Road, Chaoyang District, Beijing 100101, People's Republic of China 
   }

   \date{Received September 30, 20XX}

\abstract{Recent studies found a number of short-timescale microlensing events that are thought to be produced by free-floating or wide-orbit planets. The microlensing event KMT-2024-BLG-0792/OGLE-2024-BLG-0516 is currently the only free-floating planet candidate with a directly measured mass ($0.73^{+0.25}_{-0.15}$ Saturn masses), derived from joint ground-based and \textit{Gaia} satellite observations. However, it remains unclear whether the lens is truly isolated or bound to a widely separated host star. Here, we report high-angular-resolution interferometric observations of the event obtained with the Very Large Telescope Interferometer/GRAVITY instrument two years after peak magnification. We detect no luminous host star up to a projected separation of 90\,au, supporting the hypothesis that the lens is indeed a free-floating planet. We derive a $5\sigma$ upper limit on the mass of a putative host star of approximately $0.5\!-\!0.6\,$M$_{\odot}$. This work demonstrates that interferometry provides a powerful method for directly searching for or constraining host stars in short-duration and planetary microlensing events.}


   \maketitle

\nolinenumbers
\section{Introduction}

High-cadence observations of the Galactic bulge by modern ground-based microlensing surveys have led to the discovery of a sizable population of single, short-timescale microlensing events, which are thought to be produced by free-floating planet candidates \citep{mroz2017, gould2022, sumi2023}. These events are characterized by short Einstein timescales (\mbox{$t_{\rm E} \lesssim 1$\,d}) and small angular Einstein radii \mbox{($\theta_{\rm E} \lesssim 10\,\mu\mathrm{as}$)}, indicating that they are produced by planetary-mass objects. Because these events exhibit a singular brightening in their light curves, the lenses must be either isolated or located at wide projected separation (at least several Einstein radii) from any companion. Hence, this population of single, short-timescale microlensing events is commonly interpreted as evidence for a population of free-floating or wide-orbit planets.

It remains unclear how many of these events are produced by truly free-floating planets and how many of them have a widely separated stellar companion. \citet{yee2025} argued that all free-floating planet candidates with lenses more massive than Saturn are likely to be on wide orbits, but owing to the large host-planet separation, there is no detectable microlensing signal from the host star. The occurrence rates of wide-orbit planets (separations greater than 10\,au) more massive than Jupiter---measured from direct imaging \citep{nielsen2019,vigan2021} and radial velocity surveys \citep{fulton2021}---are consistent with the frequency of free-floating planet candidates inferred from microlensing surveys. For lower-mass wide-orbit planets in the Saturn-to-Jupiter mass range, the occurrence rate is inferred from the frequency of debris disks \citep{pearce2022}, assuming a single planet maintains the inner edge of the disk. This rate is roughly equal to the abundance of such planets in the mass function of free-floating planet candidates. Furthermore, \citet{yee2025} argued that free-floating planet events involving planetary-mass objects more massive than Jupiter are likely to orbit a massive host star ($>1\,M_{\odot}$), because such massive, wide-orbit planets are predominantly observed around stars more massive than the Sun.

The situation is less clear for lower-mass planets because current transit, radial velocity, and direct imaging experiments lack the sensitivity to detect planets smaller than Saturn on orbits wider than 10\,au. By contrast, microlensing is uniquely sensitive to this regime, and the study by \citet{poleski2021} found that microlensing stars host, on average, $1.4^{+0.9}_{-0.6}$ ice giant planets at separations from 5 to 15 au (ice giants were defined as planets with the planet-to-host mass ratio from $10^{-4}$ to 0.033 and projected separation from 2 to 6 Einstein radii). That finding prompted \citet{gould2022} to argue that wide-orbit Neptune- to Saturn-mass planets can account for the vast majority of currently known short-timescale microlensing events in that mass range.

From the theoretical perspective, \citet{hadden2026} argued that, under the hypothesis that planet--planet scattering is the dominant source of free-floating planets, roughly half of the reported ``free-floating'' Neptunes are not actually unbound, but are merely ``detached'' at separations of hundreds of au from their host stars, i.e., what \citet{gould2016} had hypothesized as ``Kuiper'' planets. Similarly, \citet{lorusso2026} studied N-body simulations of the evolution of multi-Neptune systems assembled into resonant chains during the gas-disk phase and later coupled to remnant planetesimal disks. They found that the planetesimal disk can trigger dynamical instability, producing a comparable number of bound wide-orbit planets ($>20\,\mathrm{au}$) and free-floating planets.

During a microlensing event, the lens and the source are nearly perfectly aligned (with an angular separation smaller than 0.01\,mas for typical short-timescale events) and cannot be resolved by current, or even planned, instruments. However, this situation is not static. From the observer's perspective, the lens and the source move with a typical relative proper motion on the order of 5\,mas\,yr$^{-1}$. Thus, the lens and the source will separate sufficiently on the sky to be resolved after about a year using interferometry, or about a decade using adaptive optics or space-based imaging. If the lens has a host star, its light may become detectable in high-angular resolution observations. Conversely, if the lens is a truly free-floating planet, no additional light source should be detected.

Only two published studies have attempted to determine the nature of free-floating planet candidates detected using microlensing: \citet{mroz2024e} used the Keck Telescope to observe five events with adaptive optics, and \citet{kapusta2026} analyzed serendipitous archival \textit{Hubble Space Telescope} observations of OGLE-2023-BLG-0524 taken more than 25 years prior to the event. Both efforts were ultimately inconclusive: while no stellar companions were detected, the observations lacked the depth required to definitively rule out the wide-orbit planet hypothesis.

These studies required long wait times between the event and high-resolution observations (ranging from 8 to 25 years). However, this delay can be greatly reduced using interferometric observations, which can probe the surroundings of the event at angular separations as small as a few mas. Among current optical/near-infrared interferometers, GRAVITY \citep{gravity2017} offers the highest sensitivity. It has enabled successful resolution of microlensed images in several bright events \citep[e.g.,][]{dong2019, zang2020, wu2024}.\footnote{In addition, Gaia19bld was resolved by PIONIER \citep{cassan2022}.} Moreover, the recent introduction of the dual-field wide mode \citep{gravity2022} has enabled the observation of significantly fainter events \citep{mroz2025b}. Further upgrades to the adaptive optics system have greatly improved GRAVITY's sensitivity and performance \citep{gravity2026}. As of now, about $\sim 2/3$ of all microlensing events detected by ground-based surveys can be observed with that instrument.

The microlensing event KMT-2024-BLG-0792/OGLE-2024-BLG-0516 (hereafter KB240792/OB240516) is the only known free-floating planet candidate with a directly measured mass, estimated at $0.73^{+0.25}_{-0.15}$ Saturn masses ($M_{\rm Sat}$), thanks to serendipitous observations collected by the \textit{Gaia} satellite \citep{dong2026}. We observed this event in April 2026, approximately two years after its peak, as part of the GRAVITY+ Adaptive Optics (GPAO) science verification run. Here, we present the results of a search for a putative stellar companion to the lens using high-resolution GRAVITY observations.

\section{KMT-2024-BLG-0792/OGLE-2024-BLG-0516}

The event was independently discovered by the Korea Microlensing Telescope Network (KMTNet; \citealt{kim2016}) and the Optical Gravitational Lensing Experiment (OGLE; \citealt{udalski2015}) surveys on the night of May 3, 2024. This short-timescale ($t_{\rm E} = 0.842 \pm 0.002$ days) event exhibited pronounced finite-source effects, allowing \citet{dong2026} to measure its angular Einstein radius of $\theta_{\rm E} = 18.6 \pm 0.9\,\mu\mathrm{as}$. Additionally, the event was simultaneously observed by the \textit{Gaia} satellite. The \textit{Gaia} and ground-based light curves exhibited nearly identical maximum magnifications, but due to the space-based parallax, the \textit{Gaia} light curve peaked approximately $1.9$\,hours later than the ground-based observations. By jointly modeling the ground-based and \textit{Gaia} light curves, \citet{dong2026} measured the microlensing parallax vector $\boldsymbol{\pi}_{\rm E}$, thereby determining the mass and distance of the lens.

There are four possible solutions that differ by the signs of the impact parameter \citep{refsdal1966,gould1994b}, which we denote as $(++)$, $(--)$, $(+-)$, and $(-+)$. According to \citet{dong2026}, the $(++)$ solution is preferred. In this model, the lens has a mass of $0.73^{+0.25}_{-0.15}\,M_{\rm Sat}$ and a distance of $3.05^{+0.58}_{-0.43}\,$kpc. The inferred mass and distance are similar in the $(--)$ model. Conversely, the two remaining solutions, $(+-)$ and $(-+)$, are strongly disfavored relative to the $(++)$ model by factors of $7.9 \times 10^{-8}$ and $2.4 \times 10^{-5}$, respectively \citep{dong2026}.

To predict the expected lens--source separation at the time of the GRAVITY observations, we must first calculate the relative lens--source proper motion in the barycentric frame:
\begin{equation}
\boldsymbol{\mu}^{\rm bary} = \frac{\theta_{\rm E}}{t_{\rm E}}\frac{\boldsymbol{\pi}_{\rm E}}{\pi_{\rm E}} + \frac{\boldsymbol{v}_{\oplus,\perp}\pi_{\rm rel}}{\mathrm{au}},
\end{equation}
where $\pi_{\rm rel} = \pi_{\rm E}\theta_{\rm E}$ is the relative lens--source parallax and $\boldsymbol{v}_{\oplus,\perp}=(20.58, 2.15)\,\mathrm{km}\,\mathrm{s}^{-1}$ is the velocity of the Earth in the barycentric frame at the time of the event projected onto the sky at the position of the event (along the East and North directions, respectively). The posterior vector separation $\Delta\boldsymbol{s}=(\Delta E,\Delta N)$ in East and North directions is then
\begin{equation}
\Delta\boldsymbol{s} = \boldsymbol{\mu}^{\rm bary} \Delta t,
\end{equation}
where $\Delta t=2.0$\,yr. The expected lens--source separations for the four solutions reported by \citet{dong2026} are provided in Table~\ref{tab:sep}. For the preferred $(++)$ solution, this yields an angular separation of $14.7 \pm 0.8$\,mas.

\begin{table}
\caption{Expected lens--source separation during GRAVITY observations of KB240792/OB240516.}
\label{tab:sep}
\centering
\begin{tabular}{cccc}
\hline \hline
Model & $\Delta E$ (mas) & $\Delta N$ (mas) & $\Delta s$ (mas) \\
\hline
(+ +)   & $-12.9^{+1.3}_{-1.5}$ & $7.0^{+1.6}_{-2.8}$ & $14.7 \pm 0.8$\\
(-- --) & $-14.2^{+0.9}_{-0.8}$ & $-0.6^{+2.8}_{-1.7}$ & $14.4 \pm 0.8$\\
(+ --)  & $20.9 \pm 1.0$ & $17.9 \pm 0.9$ & $27.6 \pm 1.3$\\
(-- +)  & $14.7 \pm 0.7$ & $-13.6 \pm 0.7$ & $20.0 \pm 1.0$\\
\hline
\end{tabular}
\end{table}

The fiducial models considered by \citet{dong2026} assumed no blended light, a choice supported by the location of the source star in the color--magnitude diagram and the lack of a centroid shift during the magnified portion of the light curve. Nevertheless, the best-fit models with free blending were formally preferred by $\Delta\chi^2=84$ (with $\chi^2=80833.5$ for 80,879 degrees of freedom) with $\approx32\%$ of the baseline light attributed to a blended star. While this preference was tentatively attributed to low-level systematic errors in the KMTNet data (as no significant evidence for blended light was found in the independent OGLE data), GRAVITY observations can provide a definitive, empirical resolution by directly probing the origin of this blended light.

\section{GRAVITY Observations}

We observed KB240792/OB240516 on April 30, 2026 with the GRAVITY instrument using the four 8-m Unit Telescopes (UTs) using the European Southern Observatory (ESO) Very Large Telescope Interferometer (VLTI). Observations were conducted as part of the GPAO science verification run under ESO program 116.29HF.001 (PI: P.~Mr\'oz). The baseline $K$-band and $G_{\rm RP}$ magnitudes of the target are $10.764 \pm 0.002$ \citep{minniti2010} and $13.981 \pm 0.003$ \citep{gaia_edr3}, respectively, so the source itself served as the natural guide star for the adaptive optics and the fringe-tracking reference source. The angular radius of the source is $18.4 \pm 0.9$\,$\mu$as \citep{dong2019}. The data were collected using a medium resolution mode ($R\approx 500$) in five exposures, each with the integration time $\mathrm{NDIT}\times\mathrm{DIT}=32\times 10$\,s. The atmospheric conditions were optimal with seeing in the range $0.50\!-\!0.67''$ and the coherence time $\tau_0 = 5.3\!-\!6.1$\,ms.

Data were reduced using the GRAVITY pipeline \citep{lapeyrere2014}, version 1.9.6.\footnote{https://www.eso.org/sci/software/pipelines/gravity/} A partially resolved calibrator star was observed immediately before the science target to estimate the transfer function of the instrument. The calibrator star, CD-31 15385, is a K2III giant of magnitude $K=8.65$. Based on color--surface brightness relations from \citet{kervella2004}, we estimate its angular diameter to be of $90.0 \pm 1.5$\,$\mu$as with a modest reddening of $E(B-V) \approx 0.08$ to yield the best consistency between the pairwise diameters from $V$, $J$, $H$ and $K$ bands. Because we only use closure phases, the precise value of the angular diameter is unimportant as long as the star is marginally resolved (i.e., for an angular diameter smaller than 1\,mas) and the star is not binary. The consistency of the photometric measurements with those expected from an unresolved single star, together with the fact the observed closure phases from the calibrator are only a fraction of a degree give strong indications that the calibrator can indeed be trusted. The calibrator was observed with a different exposure time than the science target, but this is taken into account by the pipeline: it estimates the phase errors due to imperfect fringe tracking during the exposure and corrects the observables, provided the option \texttt{vis-correction-sc=VFACTOR} is used (which is the default).

\begin{figure}[t]
    \centering
    \includegraphics[width=\columnwidth]{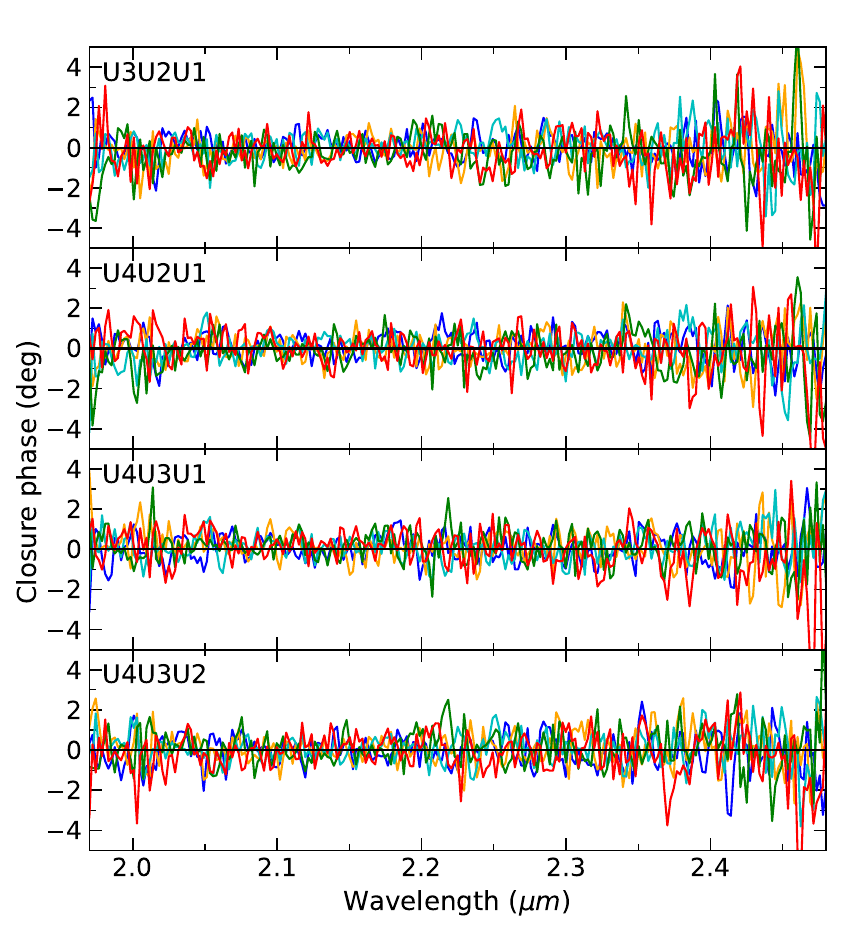}
    
    \caption{Closure phases for each triangle of UTs, with individual exposures shown in different colors.}

    \label{fig:t3phi}
\end{figure}

\section{Results}
Calibrating visibility data presents a major challenge because of atmospheric and instrumental systematic effects, which may not be fully removed during calibration, particularly when only a single calibrator is available. For this reason, we base our analysis solely on closure phases (Fig.~\ref{fig:t3phi}), which are by construction independent of atmospheric and instrumental phase errors.

In the ideal zero-noise scenario, the bispectrum of a point-symmetric source is strictly real and the corresponding closure phases are therefore either $0^\circ$ or $180^\circ$. Any deviation from these values indicates an asymmetric intensity distribution. In the context of this work, non-zero (or non-$180^{\circ}$) closure phases would indicate the presence of an off-axis companion, such as a putative host star.

To investigate whether a signature of a host star is present in our data, we implement a model-fitting strategy based on $\chi^2$ minimization. Assuming that the error bars are normally distributed and neglecting correlations between individual data points, the $\chi^2$ statistic is given by
\begin{equation}
\chi^2 = \sum_{i} \frac{(\Phi_i-\Phi_i^{\rm model})^2}{\sigma_i^2},
\label{eq:uncorr}
\end{equation}
where $(\Phi_i,\sigma_i)$ are the individual closure phases and their uncertainties, and $\Phi_i^{\rm model}$ are the closure phases predicted by a given model. To account for uncertainties introduced during the calibration process, we add 0.402\,deg in quadrature to all closure phase uncertainties reported by the pipeline. Because both the source star and a putative host star are individually unresolved, the binary-star model is fully specified by three parameters: the relative position of the host $(\Delta E,\Delta N)$ and its flux ratio $\eta$ relative to the source. In contrast, the single-star model has no free parameters and predicts closure phases equal to zero. 

Off-axis companions experience flux attenuation due to misalignment of the fiber mode center and the apparent position of the companion. To recover the companion's intrinsic flux ratio, these offset-dependent losses must be corrected a posteriori.  We account for this effect using the normalized coupling efficiency, $\gamma(\Delta E,\Delta N)$, calculated from the fundamental mode profile of the fiber following \citet{Wang_2021}. The corrected flux ratio is obtained by dividing the measured flux ratio by $\gamma$.

To determine whether a host star is detected, we find the best-fit binary-model parameters corresponding to the minimum value of $\chi^2_{\rm binary}$ and calculate the $\Delta\chi^2=\chi^2_{\rm single} - \chi^2_{\rm binary}$ difference. We consider the binary companion detected if $\Delta\chi^2 \geq \Delta\chi^2_{\rm thresh}$. Formally, if uncertainties are normally distributed and data are uncorrelated, the statistic $\Delta\chi^2$ asymptotically follows a $\chi^2$ distribution with three degrees of freedom. Achieving a $5\sigma$ significance level (corresponding to a false alarm probability of $5.73 \times 10^{-7}$) requires $\Delta\chi^2_{\rm thresh}=31.8$. However, in practice, the presence of correlated noise in the data necessitates setting larger values of $\Delta\chi^2_{\rm thresh}$.

Correlated noise in the GRAVITY closure phase data was analyzed by \citet{Kammerer_2020}, who found its two main sources: correlations between neighboring spectral channels and correlations between the same spectral channels on different telescope triangles. We model these correlations using a matrix $\boldsymbol{C}(y)$, where $y$ parameterizes the correlation length between spectral channels. The elements of the covariance matrix $\boldsymbol{\Sigma}(y)$ are defined as $\Sigma_{ij}(y) = C_{ij}(y)\sigma_i\sigma_j$, where $\sigma_i$ and $\sigma_j$ are individual closure phase uncertainties and correlations $C_{ij}(y)$ are defined as in \citet{Kammerer_2020}.
To account for non-zero diagonal terms in the covariance matrix, the objective function must be modified to
\begin{equation}
\chi^2 = (\Phi - \Phi^{\rm model})^T \boldsymbol{\Sigma}^{-1} (\Phi - \Phi^{\rm model}) + \ln|\det\boldsymbol{\Sigma}|.
\label{eq:corr}
\end{equation}

To find the best-fit value of $y$, we sample the parameter space assuming the single-star model using Markov chain Monte Carlo (MCMC) sampling with a custom implementation of the affine-invariant ensemble sampler algorithm of \citet{Goodman2010}. This yields $y = 0.016 \pm 0.007$ and throughout the remainder of this paper, we fix $y=0.016$.

\subsection{Search for a Host Star}
\label{sec:model_fitting}

We carry out a grid search for binary-star models in which a companion is placed at various relative positions $(\Delta E,\Delta N)$ with varying flux ratios $\eta$ relative to the source star.

The interferometric angular resolution is given by $\lambda/(2B_{\mathrm{max}})$, where $B_{\mathrm{max}}$ is the longest available baseline. In our case, this corresponds to approximately 1.6\,mas, setting the minimum detectable separation between the source and a potential host star. The outer boundary of the search grid is determined by the field of view of the fiber mode, which has a half width at half maximum of $0.514\lambda/D$ for an Airy function, approximately 30\,mas. This corresponds to a projected separation of about 90\,au assuming the lens distance of 3.05\,kpc. The flux ratio $\eta$ ranges from 0 to 1, with the special case $\eta=0$ representing the default single-star model.

Once a local $\chi^2$ minimum is found, we explore the parameter space using the MCMC approach. Assuming uncorrelated errors (Eq.~\ref{eq:uncorr}), we find one prominent $\chi^2$ minimum at $\Delta E=-5.48^{+0.09}_{-0.08}$\,mas, $\Delta N=3.45^{+0.18}_{-0.25}$\,mas, and $\eta=0.0068^{+0.0026}_{-0.0023}$, with $\chi^2_{\rm binary}/\mathrm{d.o.f}=4653.5/4657$. For the correlated-error model (Eq.~\ref{eq:corr}), the best-fit parameters are $\Delta E = -5.39^{+0.18}_{-0.14}$\,mas, $\Delta N = 3.16^{+0.37}_{-0.57}$\,mas, and $\eta = 0.0046^{+0.0033}_{-0.0022}$, with $\chi^2_{\rm binary} = 1678.2$.

We next evaluate the statistical significance of these binary solutions. For the single-star model, we find $\chi^2_{\rm single} = 4705.4$ (uncorrelated errors) and $\chi^2_{\rm single} = 1703.1$ (correlated error model). The corresponding $\Delta\chi^2$ values are 51.9 and 24.9, respectively. Under the uncorrelated errors model, $\Delta\chi^2=51.9$ exceeds the formal threshold of $\Delta\chi^2_{\rm thresh}=31.8$ required for a $5\sigma$ detection with three degrees of freedom. However, as discussed above, this value is artificially inflated by correlated noise that is present in the interferometric data. Once this noise is taken into account (as in our correlated error model), $\Delta\chi^2=24.9$ falls below the formal threshold. We therefore conclude that the best-fit binary-star model is not statistically significant at the $5\sigma$ level.

\subsection{Sensitivity Limits}

\begin{figure*}[!t]
    \centering
 
    \includegraphics[width=\textwidth]{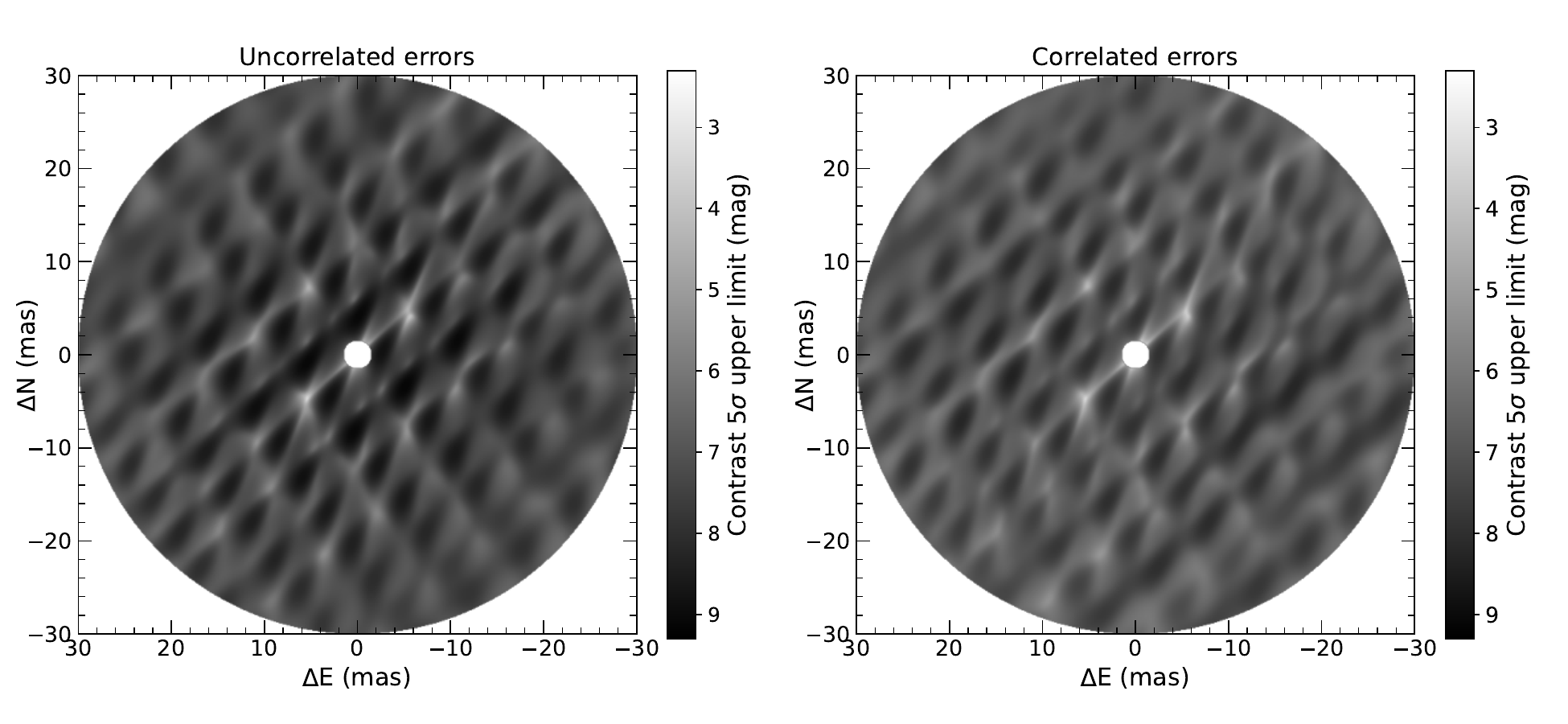}
    \caption{Comparison between $5\sigma$ flux ratio sensitivity maps obtained assuming uncorrelated errors (left panel) and the correlated error model of \citet{Kammerer_2020} (right panel). The source star is located at $(0,0)$ mas. }
    \label{fig:Smap}
\end{figure*} 

\begin{figure}[t]
    \centering
    \includegraphics[width=0.95\columnwidth]{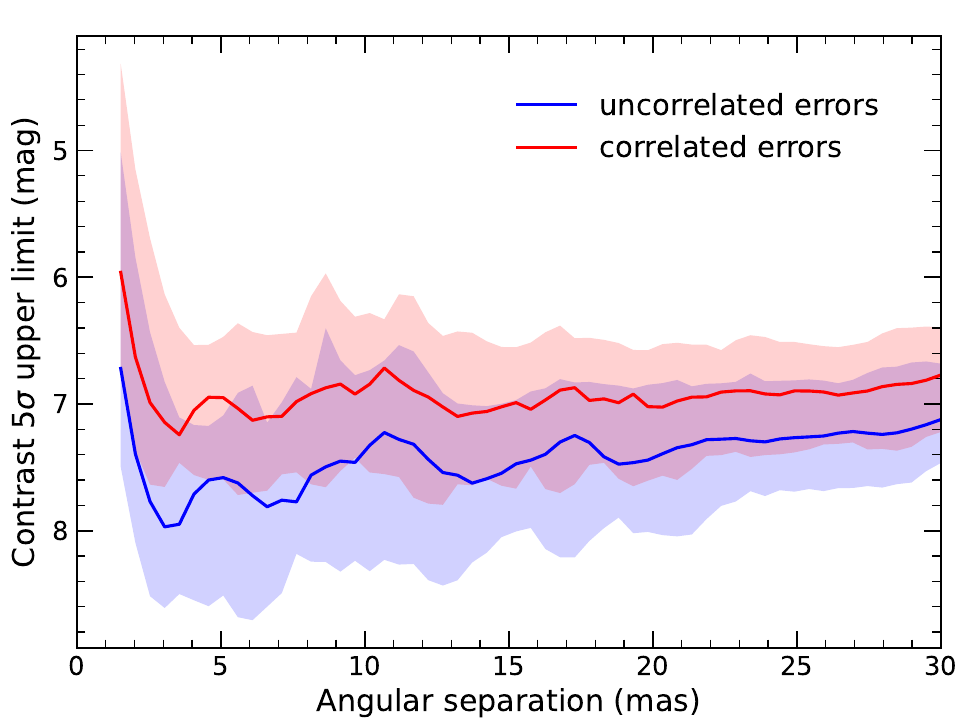}
    
    \caption{Comparison of contrast curves obtained by computing the azimuthal median flux ratio as a function of angular separation. The shaded bands represent the 68\% percentile interval of the flux ratios at each separation. }

    \label{fig:contrast_mag}
\end{figure}

To assess the companion flux sensitivity limits across GRAVITY's field of view, we use the fact that, for a faint companion ($\eta \ll 1$), the closure phases $\Phi$ scale linearly with the flux ratio (see \citealt{Kammerer_2019,Kammerer_2020}):
\begin{equation}
\Phi\left(\eta\right) \propto  \eta.
\end{equation}
(In Appendix~\ref{sec:appendix}, we discuss the validity of this assumption.)

We define $M_{\mathrm{ref}}$ as the ratio between the closure phase signal of a reference faint companion and its flux ratio $\eta_{\mathrm{ref}}$, $M_{\mathrm{ref}}= {\Phi\left(\eta_{\mathrm{ref}}\right)}/{\eta_{\mathrm{ref}}},$ where $\eta_{\mathrm{ref}}= 10^{-5}$. The analytical best-fit flux ratio $\eta_{\rm fit}$ and its uncertainty $\sigma_{\rm fit}$ at a given grid position are given by

\begin{align}
\eta_{\mathrm{fit}} &= \frac{M_{\mathrm{ref}}^T \, \boldsymbol{\Sigma}^{-1} \, D}{M_{\mathrm{ref}}^T \, \boldsymbol{\Sigma}^{-1} \, M_{\mathrm{ref}}}, \label{eta1} \\
\sigma_{\mathrm{fit}} &= \frac{1}{\sqrt{ M_{\mathrm{ref}}^T \, \boldsymbol{\Sigma}^{-1} \, M_{\mathrm{ref}}}},
\label{sgm1}
\end{align}
where $D$ denotes the closure phase data vector.

Under the assumption that no host star is present, the $5\sigma$ flux ratio limit is nominally $\eta_{\mathrm{lim}} = \eta_{\mathrm{fit}} + 5\sigma_{\mathrm{fit}}$. However, due to statistical fluctuations, we obtain $\eta_{\mathrm{lim}}<0$ or $\eta_{\mathrm{fit}}<0$ in some grid points. To restrict the solution to positive values, we adopt the following expression for the $5 \sigma$ flux ratio limit
\begin{equation}\eta_{\mathrm{lim}}=\sqrt{2} \sigma_{\mathrm{fit}} \, \mathrm{erf}^{-1}\left(1- P_0\left(1-\mathrm{erf}\left(\frac{-\eta_{\mathrm{fit}}}{\sqrt{2} \sigma_{\mathrm{fit}}}\right)\right)\right) + \eta_{\mathrm{fit}},
\label{eta_erf}
\end{equation}
where $P_0 = 5.73 \times 10^{-7}$. This formulation is mathematically equivalent to requiring $\chi^2(\eta=\eta_{\rm lim})=\chi^2_{\rm min} + 25$, where $\chi^2_{\rm min}$ is the minimal $\chi^2$ at a given position and 25 is the critical value of a $\chi^2$ distribution with one degree of freedom corresponding to a significance level of $5.73 \times 10^{-7}$.

Fig.~\ref{fig:Smap} compares the resulting 2D sensitivity maps derived assuming uncorrelated errors (left panel) and the correlated error model of \citet[][right panel]{Kammerer_2020}. The repetitive pattern present in both maps reflects the sparse $uv$-plane coverage, which causes the sensitivity to vary periodically over the field of view. Additionally, we derive the corresponding contrast curves (Fig.~\ref{fig:contrast_mag}) by computing the azimuthal median of the $5\sigma$ flux ratio upper limits as a function of angular separation from the source star. We observe that the $5 \sigma$ flux ratio upper limit decreases on average up to approximately 3.5\,mas from the source, after which it remains relatively constant, then increases again toward the edge of the field of view due to fiber injection losses. The slope of the contrast curve for the correlated error model (red curve)  is shallower than that for the uncorrelated error model (blue curve). This behavior results from two competing effects: while off-axis fiber injection losses degrade sensitivity at wider separations, spatial noise correlations simultaneously get weaker, slightly improving the relative flux limits. 
Across all separations, we observe a difference in the sensitivity of two methods, with the correlated error model being approximately 0.3--0.9\,mag less sensitive than the limits calculated assuming uncorrelated errors. This conclusion contrasts with that of \citet{Kammerer_2020}, which we discuss in more detail in Appendix~\ref{sec:absil}.


\section{Limits on a Host Star}

\begin{figure*}[t]
    \centering
    \includegraphics[width=\textwidth]{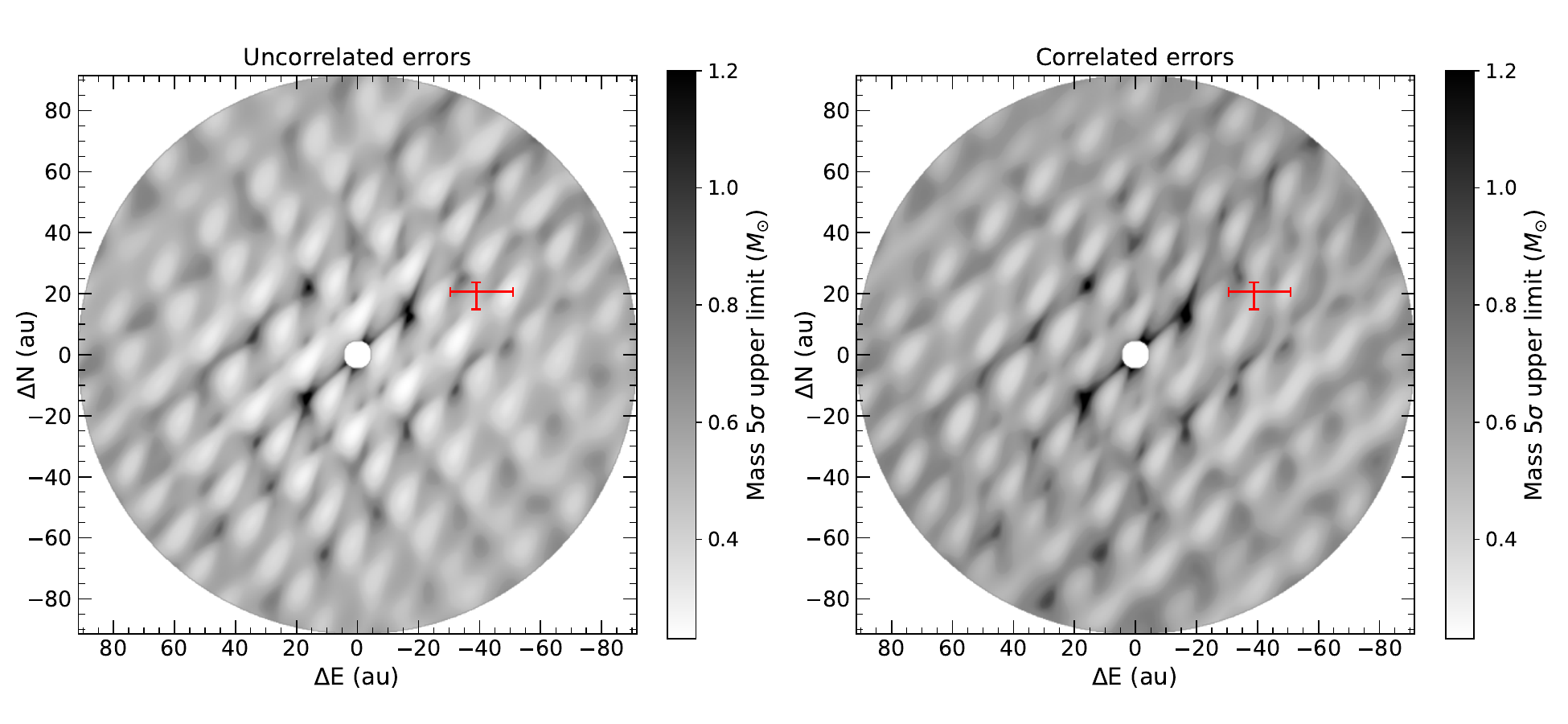}
    \caption{Comparison between $5\sigma$ host star mass sensitivity maps calculated from the flux ratio limits found assuming uncorrelated errors (left panel) and the correlated error model of \citet{Kammerer_2020} (right panel). The source star is located at $(0,0)$\,mas. The predicted position of the lens in the $(++)$ solution of \citet{dong2026} is marked by a red point. }
    \label{fig:Smap_mass}
\end{figure*}

We convert sensitivity limits on the flux ratio into upper limits on the mass of the host star. To this end, we determine the absolute $K_s$ magnitude of the host star using 
 \begin{equation}
     M_{K_{s}}= m_{K_{s}}- 5 \log(D)-A_{K_{s}}+5,
 \end{equation}
where $m_{K_s}$ is the apparent $K_s$ magnitude of the host star computed from the $5 \sigma$ flux ratio sensitivity limits and the apparent magnitude of the source, $D$ is the estimated distance to the lens, and $A_{K_s}$ is the extinction along the line of sight to the lens. We fix $D=3.05^{+0.58}_{-0.43}$ kpc, as given by the $(++)$ solution of \citet{dong2026}. We approximate the extinction at the lens distance by assuming that it is proportional to the dust column density along the line of sight. Therefore, we integrate the dust density along the line of sight to both the lens and the source (located in the  Galactic bulge). We model the dust density as
\begin{equation}
    \rho(R,z)= \rho_0 \exp\left(-\frac{R}{h_R}\right) \exp\left(- \frac{|z|}{h_z}\right),
\end{equation} 
where $\rho_0$ is the normalization, $R$ and $z$ are the distances to the Galactic center and Galactic plane, respectively, and $h_R=4.2$\,kpc and $h_z=88$\,pc are parameters describing the dust geometry \citep{Sharma_2011}. From these numerical calculations, we obtain an extinction toward the lens of $A_{K_s}^{\mathrm{lens}}=0.53 \, A_{Ks}^{\mathrm{source}}$. To compute the extinction toward the source, $A_{K_s}^{\mathrm{source}}$, we adopt a color excess of $E(J-K_s)= 0.355$ from the red clump stars reddening map of \citet{Surot_2020}, evaluated in the direction of the event. Additionally, we adopt a selective-to-total extinction ratio in the $K_s$ band for the Galactic bulge from \citet{Alonso_Garc_a_2017}, i.e. $A_{K_s} /E(J - K_s) = 0.428$.

\begin{figure}[t]
    \centering
    \includegraphics[width=0.95\columnwidth]{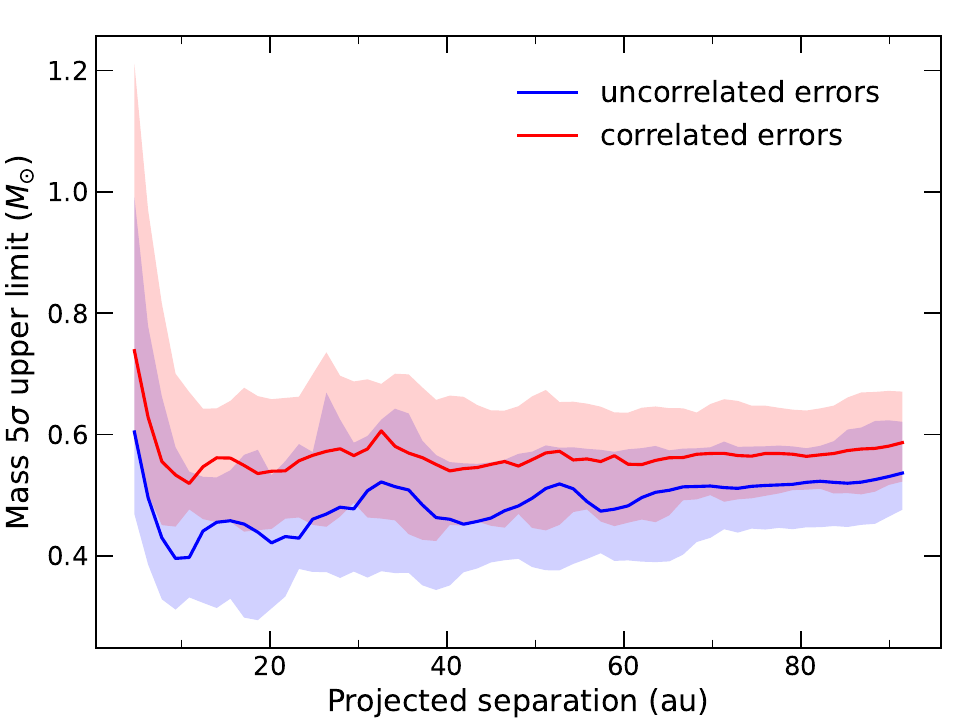}
    \caption{Comparison of host star mass limits obtained by azimuthal median of the 2D mass sensitivity maps as a function of angular separation.}
    \label{fig:contrast_mass}
\end{figure}

Once we obtain the limits on the absolute magnitude, we adopt the empirical mass--absolute magnitude relation from \citet{Pecaut_2013}, which was derived from average properties of main-sequence stars, to convert magnitude limits into mass limits. This yields mass sensitivity maps and mass contrast curves for both the correlated and uncorrelated error methods (Figs.~\ref{fig:Smap_mass} and~\ref{fig:contrast_mass}). We observe that the $5\sigma$ upper mass limit decreases with separation up to approximately 10\,au, beyond which it remains relatively constant before increasing slightly again at larger separations. The sensitivity difference between the two methods remains, with the correlated-error method of \citet{Kammerer_2020} yielding a mass limit approximately 0.1\,M$_{\odot}$ higher than that obtained assuming uncorrelated errors. 

The adopted empirical relations provide only approximate host star masses as they are derived for a specific stellar population, whereas the nature of the potential host star remains unknown. Therefore, we complement this analysis using theoretical stellar models. Specifically, we use PARSEC stellar evolution models \citep{Bressan_2012} to derive the host star mass limits. The resulting values are in good agreement, with a percentage difference between the azimuthal medians obtained with two relations of less than 17.0\%  (15.1\%) at any angular separation  and on average equal to  10.8\%   (8.2\%) when using the Gaussian error approach (and correlated error approach).

\section{Conclusions}

Among known free-floating planet candidates, KB240792/OB240516 is unique because simultaneous ground-based and \textit{Gaia} observations enabled \citet{dong2026} to determine the microlensing parallax of the event and, as a result, both the mass of and distance to the lens. In addition, these measurements allow us to compute the relative lens--source proper motion and predict the expected position of the lens at the epoch of our GRAVITY observations. For the favored $(++)$ and $(--)$ solutions, the predicted lens locations relative to the source are $(\Delta E,\Delta N) = (-12.9^{+1.3}_{-1.5},7.0^{+1.6}_{-2.8})$\,mas and $(-14.2^{+0.9}_{-0.8},-0.6^{+2.8}_{-1.7})$\,mas, respectively. At these specific locations, we rule out any main-sequence star more massive than 0.47\,M$_\odot$ and 0.41\,M$_{\odot}$ (0.47\,M$_{\odot}$ and 0.51\,M$_{\odot}$) assuming that closure phase errors are uncorrelated (correlated).

Because the lens in this case is a Saturn-mass planet, any putative host star may be located at an arbitrary position relative to the lens. From the analysis of the light curve of the event, we know that a host star must be located at a projected separation greater than $27\,\mathrm{au}\sqrt{M_{\rm host}/M_{\odot}}$, where $M_{\rm host}$ is the host mass \citep{dong2026}. Otherwise, a second, long-timescale brightening would have been detected in the event's light curve. 

Therefore, the upper limits on the mass of a potential host star depend on its sky position, as presented in Fig.~\ref{fig:Smap_mass}. Using the uncorrelated-error approach, we typically rule out main-sequence host stars more massive than 0.4\,M$_{\odot}$ at a projected separation of 10\,au from the source, with the limits weakening to about 0.55\,M$_{\odot}$ at 90\,au (Fig.~\ref{fig:contrast_mass}). When accounting for correlated noise that is present in the GRAVITY data following \citet{Kammerer_2020}, the limits become less stringent: from about 0.5\,M$_\odot$ at 10\,au to 0.6\,M$_{\odot}$ at 90\,au (Fig.~\ref{fig:contrast_mass}).

This work also demonstrates that interferometric observations provide a powerful method for directly searching for or constraining host stars in free-floating planet microlensing events. In contrast to direct imaging, which requires long delay times ($\gtrsim 10$\,years) between the event and follow-up observations \citep[e.g.,][]{mroz2024e,kapusta2026}, interferometry allows us to probe the surroundings of an event at milliarcsecond scales as early as a year after the event. Long delay times are actually problematic because high proper motion lenses can quickly exit the 60\,mas field of view of GRAVITY. Another promising free-floating planet candidate suitable for GRAVITY follow-up is KMT-2024-BLG-0816/OGLE-2024-BLG-0519 \citep{poleski2025}. Because its relative lens--source proper motion is smaller than that of KB240792/OB240516, we predict a lens--source separation of approximately 12\,mas in 2027.

Finally, GRAVITY can also be applied to study bound exoplanets that are detected via gravitational microlensing. In these systems, where the existence of the host star is guaranteed, measuring the flux and position of the host relative to the source directly breaks the mass--distance degeneracy that inherently limits many microlensing discoveries.

\begin{acknowledgements}
We thank Prof.~Andrew Gould for his comments on the manuscript.
Based on observations collected at the European Southern Observatory under ESO program 116.29HF.001. This research was funded in part by National Science Centre, Poland, grant SONATA 2023/51/D/ST9/00187 awarded to P.M.
\end{acknowledgements}

\bibliographystyle{aa} 
\bibliography{pap} 


\begin{appendix}

\nolinenumbers

\section{Testing the Linearity between Closure Phase and Faint Companion Flux}
\label{sec:appendix}

Equations (\ref{eta1}) and (\ref{sgm1}) were derived by minimizing the $\chi^2$ function, under the assumption that closure phases depend linearly on the flux ratio of a faint companion ($\eta \ll 1$). Here, we test the validity of this linearity assumption through an independent numerical check.

To this end, we perform MCMC sampling at several fixed spatial positions assuming the correlated-error model of \citet{Kammerer_2020} (Eq.~\ref{eq:corr}). Since we have no prior knowledge of the correlations in our data, we jointly sample over the flux ratio $\eta$ and the correlation parameter $y$. To ensure physically meaningful solutions, we impose a flat prior enforcing $\eta \geq 0$ during sampling. 

\begin{figure}[t]
    \includegraphics[width=0.5\textwidth]{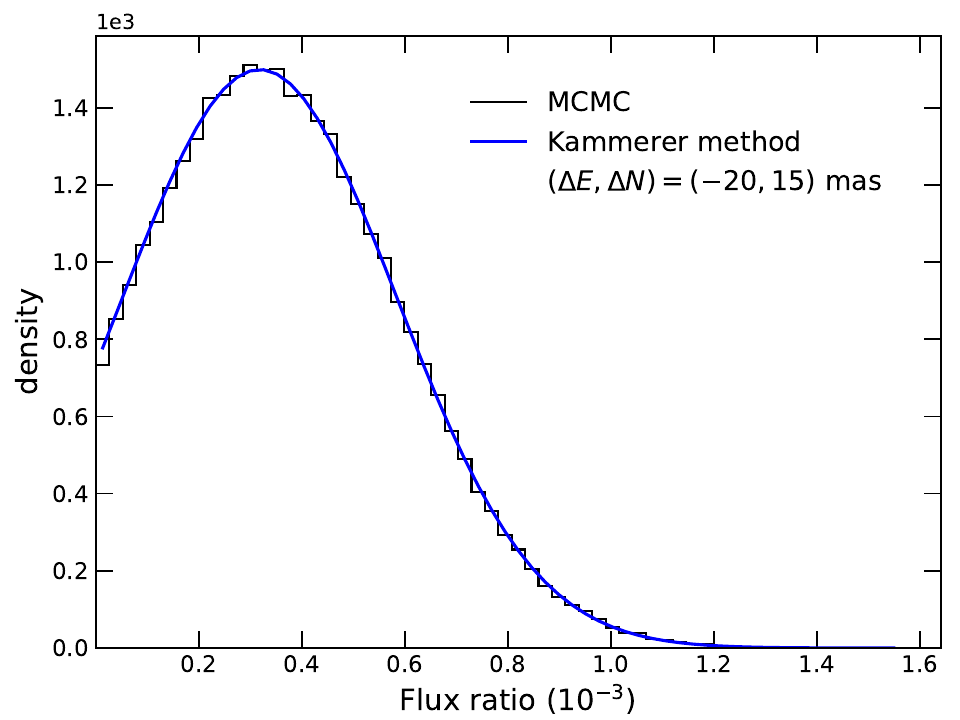}
    \includegraphics[width=0.5\textwidth]{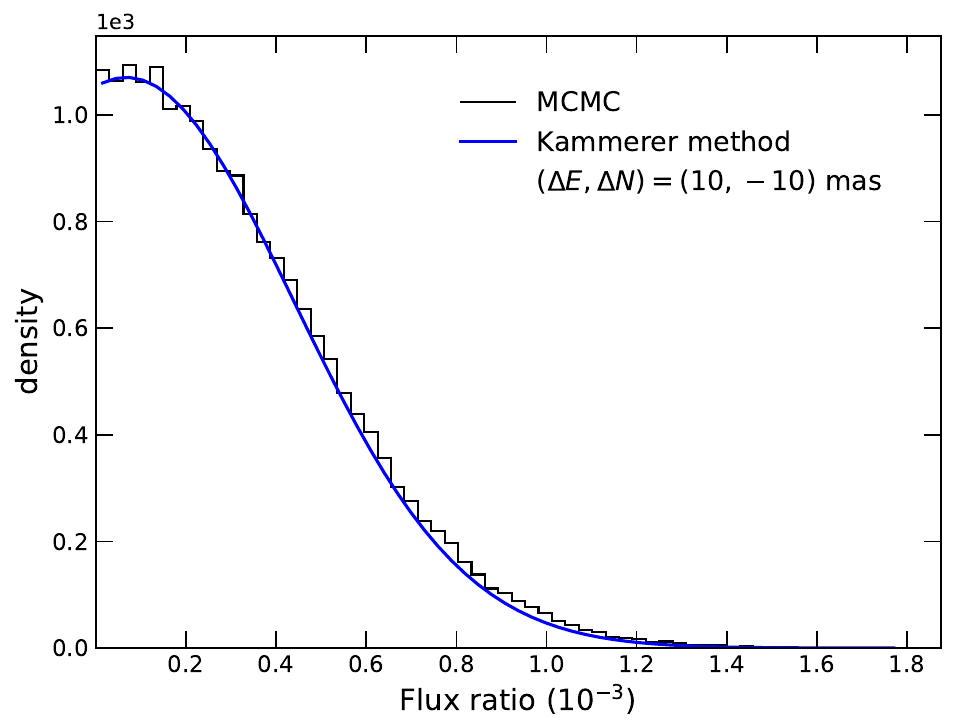}
    \caption{Comparison between the posterior distributions of flux ratio $\eta$ obtained from MCMC (black histogram) and the analytical Gaussian distribution with mean $\eta_{\rm fit}$ and standard deviation $\sigma_{\rm fit}$ (blue line) derived following \citet{Kammerer_2020}. Left panel: $(\Delta E, \Delta N)=(-20,15)$\,mas, right panel: $(\Delta E, \Delta N)=(10,-10)$\,mas.}
    \label{fig:Eta_dist}
\end{figure}

We then compare the resulting MCMC posterior distribution for $\eta$ with a Gaussian distribution defined by the analytical mean $\eta_{\mathrm{fit}}$ and standard deviation $\sigma_{\mathrm{fit}}$ computed from Eqs.~(\ref{eta1}) and~(\ref{sgm1}). The parameter distribution is expected to be Gaussian under the assumption that the $\chi^2$ function is smooth near its minimum.
After examining several representative locations in the field of view (Fig.~\ref{fig:Eta_dist}), we conclude that the MCMC posterior distributions agree well with the analytical Gaussian models. This confirms that the linear approximation used to derive $\eta_{\mathrm{ fit}}$ and $\sigma_{\mathrm{fit}}$ is sufficiently accurate for our sensitivity calculations. Consequently, we can reliably use these analytical expressions to evaluate $5\sigma$ sensitivity limits, rather than running MCMC sampling at every grid point, which would require more computational resources.

\section{Sensitivity Limits from the \citet{Absil_2011} Method}
\label{sec:absil}

Another method commonly used to derive upper flux ratio limits for binary companions in interferometric data was introduced by \citet{Absil_2011}. In this framework, one calculates the probability
\begin{equation}
    P(\Delta E,\Delta N, \eta) = 1 - \mathrm{CDF}_\nu\left(\frac{\chi^2_{\rm binary}(\Delta E,\Delta N,\eta)}{\chi^2_{\rm single}/{\nu_1}}\right),
    \label{eq:absil}
\end{equation}
where $\nu$ and $\nu_1$ are the degrees of freedom for the binary-star and single-star models, respectively, and $\mathrm{CDF}_\nu$ is the cumulative distribution function of a $\chi^2$ distribution with $\nu$ degrees of freedom. One then determines the limiting flux ratio $\eta_{\rm lim}$ such that $P(\Delta E,\Delta N,\eta_{\rm lim})=P_0=5.73 \times 10^{-7}$ for a $5\sigma$ significance limit. The expression in the denominator $\chi^2_{\rm single}/{\nu_1}$ effectively serves to normalize the observational uncertainties (by rescaling the reduced $\chi^2$ of the single-star model to unity) rather than acting as a relative model-comparison statistic. Consequently, $P(\Delta E,\Delta N, \eta)$ evaluates the absolute goodness of fit of the binary-star model to the rescaled data, rather than testing whether the binary-star model is statistically preferred over the single-star model.

As a result, the upper limits obtained using Eq.~(\ref{eq:absil}) require a significantly larger change in $\chi^2$ than those derived from a standard $\Delta\chi^2$ model-selection statistic. For example, with $\nu = 4659$ degrees of freedom in our dataset, achieving $P=P_0$ requires a $\chi^2$ change by $\Delta\chi^2=\chi^2_{\rm single}-\chi^2_{\rm binary} \approx 485$, which corresponds to a nearly $22\sigma$ significance limit under a standard $\Delta\chi^2$ statistic.

When discussing the contrast limits achievable when accounting for the correlations in the interferometric data, \citet{Kammerer_2020} concluded that faint source detection limits improve by up a factor of $\sim\! 2$ compared to the method of \citet{Absil_2011}. However, this apparent gain occurs because the two methods require fundamentally different $\Delta\chi^2$ detection thresholds. In fact, as shown in Fig.~\ref{fig:contrast_mag}, accounting for correlated noise in the data actually yields less stringent contrast limits than when assuming uncorrelated errors (when evaluating both under the same significance threshold). This is a natural consequence of correlated noise, which makes faint off-axis companions harder to detect.

\end{appendix}

\end{document}